\documentclass[prd,showpacs,preprintnumbers,amsmath,amssymb,nofootinbib]{revtex4}

\usepackage{graphicx}% Include figure files
\usepackage{dcolumn}% Align table columns on decimal point
\usepackage{bm}% bold math
\usepackage{dsfont}

\newcommand{\Od}{{\cal O}}

\newcommand{\Tr}{\mbox{Tr}}

\newcommand{\re}{\mbox{Re}\,}

\newcommand{\diag}{\mbox{diag}}

\newcommand{\condtwo}{\langle \bar q q \rangle}
\newcommand{\conds}{\langle \bar s s \rangle}
\newcommand{\condfour}{\langle (\bar q q)^2 \rangle}
\newcommand{\condn}{\langle (\bar q q)^n \rangle}
\newcommand{\quarkcor}{\langle T (\bar q q)(x) (\bar q
q)(0)\rangle}
\newcommand{\quarkcorl}{\langle T (\bar q q)_l(x) (\bar q
q)_l(0)\rangle}

\newcommand{\be}{\begin{equation}}
\newcommand{\ee}{\end{equation}}
\newcommand{\ba}{\begin{eqnarray}}
\newcommand{\ea}{\end{eqnarray}}

\newcommand{\gsim}{\raise.3ex\hbox{$>$\kern-.75em\lower1ex\hbox{$\sim$}}}
\newcommand{\lsim}{\raise.3ex\hbox{$<$\kern-.75em\lower1ex\hbox{$\sim$}}}

\begin{document}
%\baselineskip=20pt
% declarations for front matter

\title{
Non-factorization of four-quark condensates
at low energies within Chiral Perturbation Theory
 }

\author{A. G\'omez Nicola}
\email{gomez@fis.ucm.es} \affiliation{Departamento de F\'{\i}sica
Te\'orica II. Universidad Complutense. 28040 Madrid. Spain.}
\author{J.R. Pel\'aez}
\email{jrpelaez@fis.ucm.es} \affiliation{Departamento de
F\'{\i}sica Te\'orica II. Universidad Complutense. 28040 Madrid. Spain.}
\author{J. Ruiz de Elvira}\email{jacobore@rect.ucm.es}
\affiliation{Departamento de F\'{\i}sica Te\'orica II. Universidad
Complutense. 28040 Madrid. Spain.}

\begin{abstract}
Four-quark correlators and the factorization hypothesis are
analyzed in the meson sector within
Chiral Perturbation Theory. We define the four-quark condensate as  $\lim_{x\to 0}\quarkcor$, which is equivalent to other definitions  commonly used in the literature. Factorization of the four-quark condensate holds to
leading and next to leading order. However
at next to next to leading order, a term with a nontrivial space-time dependence in the
four-quark correlator yields a divergent four-quark condensate,
whereas the two-quark condensate and the scalar susceptibility are finite.
Such a non-factorization term vanishes only in the chiral limit.
We also comment on how factorization still holds in the large $N_c$ limit,
provided such a limit is taken before renormalization.
\end{abstract}

\pacs{12.39.Fe, 11.30.Rd, 11.15.Pg}

%\vspace{-.5cm}
%\rule{\textwidth}{.1mm}

\maketitle

\section{Introduction}

Scalar condensates play a relevant role in QCD, since they are
directly related to  vacuum properties. The quark condensate
$\condtwo$ is a parameter deeply related to spontaneous chiral symmetry
breaking and the description of
low-energy QCD. In principle, quark condensates of arbitrary order
$\condn$ are also built out of chiral noninvariant operators with
the vacuum quantum numbers and are also related to chiral symmetry restoration.
In addition,  quark condensates appear directly in QCD sum
rules, through the  operator product expansion
(OPE) approach \cite{Shifman:1978bx},
where the following hypothesis of factorization or
vacuum saturation is customarily made:
\begin{equation}
\condfour=\left(1-\frac{1}{N}\right)\condtwo^2.
\label{facthyp}
\end{equation}
Note that we have particularized to the case where the four-quark operator has
the quantum numbers of  the scalar, isoscalar, colorless condensates that we
are interested in. In addition, $N=4N_c N_f$, where $N_c$ and $N_f$ denote the
number of colors and flavors respectively and $q$ is a Dirac spinor,
flavor and color vector. We remark that in the large-$N_c$
limit factorization simply reduces to $\condfour=\condtwo^2$. The
second term in Eq.(\ref{facthyp}) comes from the contraction of indices (including color)
between the first and second $\bar q q$ operators.

The use of the factorization hypothesis is a key point in order
to estimate the size of higher order condensates in the OPE.
However, its justification is still a matter of debate. It was shown in \cite{Narison:1983kn} that factorization implies that $\condfour$ becomes dependent on the
QCD renormalization scale. This means that for QCD sum rules including six-dimensional operators, like $(\bar q q)^2$,
one cannot write a  renormalization-group (RG) invariant four-quark condensate, preventing RG improvements of such sum rules.
This is not a problem when considering six-dimensional pure-gluon operators or quark operators with dimensions
 lower than six, like the RG-invariant
 $ \bar q {\cal M} q$  with ${\cal M}$ the
mass matrix.  We will come back to this point in section \ref{sec:fact}. The validity of vacuum saturation has also been
questioned within the framework of finite-energy sum-rules
\cite{Bertlmann:1987ty} and has been formally
shown not to hold  when  dressed QCD vertices are considered
\cite{Zong:2006dw}.

In  this work we will present a study of the scalar
four-quark condensate
within the framework of Chiral Perturbation
Theory (ChPT). Since ChPT relies only on symmetries
and not on vacuum saturation or dominance assumptions, as in some of the approaches commented above,
it will
allow us to obtain low-energy model-independent results concerning the factorization hypothesis.

An important point concerns the definition of the quark condensate in terms of Green functions. In the chiral lagrangian framework,
 one has not access to individual quark operators at a given space-time point $x$, but to the low-energy representation of the quark-antiquark  operator $\bar q q (x)$, given by a functional derivative with respect to an external scalar source (see details in section \ref{sec:fourq}). Therefore,  a natural way to define the four-quark condensate is through the limit of the two-point function (four-quark correlator):

 \begin{equation}
 \condfour=\lim_{x\to 0}\quarkcor
 \label{ourdef4q}
 \end{equation}

This is the definition that we will choose to work with here, where all the divergencies will be treated within the $\overline{MS}$ scheme in dimensional regularization, as it is customary in ChPT. However, from the comments above, it is not clear that the four-quark condensate itself has to be a scale-independent and finite object, which means that the $x\to 0$ limit is ill-defined and  other definitions in terms of Green functions could give different answers. Actually,  Eq.(\ref{ourdef4q}) is not the usual $\overline{MS}$ definition when working for instance with four-quark vacuum expectation values in the context of electroweak penguin contributions \cite{Bijnens:2001ps,Cirigliano:2002jy}, where the following prescription is used instead:

\begin{equation}
\condfour= \int d^D x \ \quarkcor \delta^{(D)}(x)=\int\frac{d^D Q}{(2\pi)^D} \Pi (Q^2)
\label{usualdef4q}
\end{equation}
where the integrals are defined in  {\em Euclidean} space-time dimension $D$ and $\Pi(Q^2)$ is the Fourier transform of the correlator $\quarkcor$.
 In the ChPT framework, we will show (details are given in Appendix \ref{app:usualmsbar}) that this definition gives the same result as the one in Eq.(\ref{ourdef4q}) meaning that factorization is spoiled at next to next to leading order (NNLO), which  questions seriously the validity of the factorization hypothesis, now from the point of view of the low-energy representation.

The four-quark two-point
correlator, apart from defining the four-quark
condensate,  is also related the chiral or scalar susceptibility,
defined as $\chi=-\partial \condtwo /\partial m_q$ and which can be written also in terms of $\quarkcor$. The susceptibility  is a
crucial observable regarding chiral symmetry restoration, since
it is associated to thermal fluctuations and tends to grow near the
critical point \cite{Karsch:2008ch}. For us, the susceptibility will serve as a crucial consistency check, since we can calculate it directly as a quark mass derivative or through the four-quark correlator and both should coincide and be finite and scale independent.

Therefore, we will give the complete results in ChPT for the four-quark correlators and four-quark condensates in $SU(2)$ and $SU(3)$
up to NNLO, performing a consistency check by calculating the scalar susceptibility and showing the robustness of the result under different definitions of the vacuum four-quark expectation value. In addition, the discussion of factorization breaking  necessarily implies the calculation and renormalization of the two-quark condensate also at NNLO, which we will perform also explicitly here. We will also carry out the large-$N_c$ analysis of the factorization breaking, which can also be performed from the low-energy representation and is formally relevant. These are the main results of this work.

 The plan of the paper is the following: In section \ref{sec:fourq}
we present our calculation of the relevant four-quark correlators for two and three flavors.
The details of the calculation are given for $N_f=2$ for simplicity. The scalar susceptibility derived from the four-quark condensate is
 obtained in section \ref{sec:sus}.
The factorization hypothesis is then examined in section \ref{sec:fact},
whereas in section \ref{sec:largen} we discuss the large-$N_c$ limit of our results, regarding factorization.
In section \ref{sec:conclusions}, we present a brief summary and our conclusions.
Finally, in Appendix \ref{app:renquark} we provide the detailed mathematical expressions
 for the two-quark condensates to NNLO in ChPT and discuss in detail their renormalization, whereas in Appendix \ref{app:usualmsbar} we show the equivalence of our definition of the four-quark condensate with the usual one in the literature.

\section{Four-quark correlators}
\label{sec:fourq}

Our main object of study will be the time-ordered four-quark correlator $\quarkcor$.
We will follow the external source method and write this four-quark correlator as a second functional derivative of the QCD generating functional $Z_{QCD}[s]$ with respect to the scalar source $s(x)$, which in general will be a matrix-valued function in flavor space and couples to the QCD Lagrangian as:
\begin{eqnarray}
Z_{QCD}[s]&=&\int {\cal D}\bar q {\cal D}q \ldots \exp{i\int d^4 x {\cal L}_{QCD}[\bar q,q,s(x),\ldots]},
\nonumber\\
{\cal L}_{QCD}[s]&=&\bar q\left(i\not \! \! D-s(x)\right)q+\ldots,
\label{zqcd}
\end{eqnarray}
where the rest of the  Lagrangian and fields indicated by dots are  irrelevant for our purposes
and sum over $N_f$ light flavors, $N_c$ colors and Dirac indices is assumed in $\bar q q$. The physical QCD Lagrangian and partition function correspond to setting $s(x)={\cal M}$, the quark mass matrix, in the above equation.

We will consider the effective low-energy representation of $Z_{QCD}[s]$ given by Chiral Perturbation Theory \cite{Gasser:1984gg},  built from chiral symmetry invariance as an expansion in external momenta (derivatives) and meson masses:
\begin{eqnarray}
Z_{QCD}[s]&\simeq&Z_{eff}[s]=\int {\cal D}\phi^a \exp{i\int d^4 x {\cal L}_{eff}[\phi^a,s(x)]},
\nonumber\\
{\cal L}_{eff}&=&{\cal L}_2+{\cal L}_4+{\cal L}_6\ldots,
\label{zeff}
\end{eqnarray}
where the subscript in the effective Lagrangian indicate the order in the
derivative and mass expansion, formally  ${\cal L}_k={\cal O}(p^k)$
($s=\Od(p^2)$ in the standard ChPT power counting). Note that $\phi^a$ denote the NGB fields, usually collected in the $SU(N_f)$ matrix $U=\exp [i\lambda_a\phi^a/F]$, where $\lambda_a$ are the Gell-Mann or Pauli matrices for $N_f=3$ and $N_f=2$, respectively and $F$ is the pion decay constant in the chiral limit.  The Lagrangian ${\cal L}_2$ is  the non-linear sigma model:
\begin{equation}
{\cal L}_2=\frac{F^2}{4}\Tr\left[\partial_\mu U^\dagger\partial^\mu U+\chi\left( U+U^\dagger\right) \right],
\label{l2}
\end{equation}
with $\chi=2B_0 s(x)$. When $s(x)={\cal M}$, the constants $m_q,F,B_0$ appearing in ${\cal
  L}_2$ are related to meson masses, decay constants and to the quark condensate.
For simplicity, we will work in the isospin limit $m_u=m_d\equiv m$, so that,
to lowest order in $SU(2)$, $M_{0\pi}^2=2mB_0(1+\Od(p^2))$, $F_\pi=F(1+\Od(p^2))$ and
$\condtwo=B_0 F(1+\Od(p^2))$. As usual $M_{0\pi,0K,0\eta}$
stand for the leading order meson masses, in terms of which we will
express our results. Their relation to the physical masses is given in Eqs.\eqref{physicalmassesSU2} and \eqref{physicalmassesSU3}
in the Appendix \ref{app:renquark}. In addition, and for our purposes here, Weinberg's chiral power counting \cite{we79}, on which Chiral Perturbation Theory relies, can be equivalently accounted for   by keeping trace of  inverse powers of $F$, which  will be used extensively along this work.

The Lagrangians ${\cal L}_4$ and ${\cal L}_6$ are given in \cite{Gasser:1984gg}
and \cite{Bijnens:1999sh}, respectively, where use has been made of
different operator identities, partial integration and the equations of
 motion to the relevant order. Those lagrangians  contain the so-called
low-energy constants (LEC) multiplying each of the independent terms compatible
with the symmetries.  The ${\cal L}_4$ LEC receive
different names
depending on whether they multiply
terms containing $U$ fields or not; respectively, $L_i$ and $H_i$ in the SU(3) case.
The terms
without $U$ fields are contact terms containing
just external sources and no fields, but they are
needed to absorb some divergences coming from loop diagrams using ${\cal L}_2$ vertices.
The original SU(2) lagrangians in \cite{Gasser:1983yg} are written in terms of
vector fields instead of matrix fields $U$ as above, but they also use different
names for the ${\cal L}_4$ low-energy constants -- $l_i$ and $h_i$ in this case.
However, it is possible to recast \cite{Scherer:2002tk} these lagrangians using matrix field notation,
that we will use throughout this paper, and keep the same $l_i,h_i$ low energy
constants.  The relation between the SU(3) and SU(2)  low-energy constants is given in
\cite{Gasser:1984gg}, \cite{Gasser:2007sg} and
\cite{Gasser:2009hr}.

This name differentiation for the ${\cal L}_6$ is not followed any longer
\cite{Bijnens:1999sh}:
all of them are called $c_i$ in the SU(2) case and $C_i$ in the SU(3) case.
Note that the $\Od(p^6)$ LEC contained in ${\cal L}_6$ absorb
both two-loop divergences from ${\cal L}_2$ and one-loop
divergences in diagrams with  ${\cal L}_4$ vertices. All the  details for renormalization
of quark condensates up to the order we are considering here are given in Appendix \ref{app:renquark}. We recall  that the ${\cal L}_4$ Lagrangian in $SU(3)$ contains also the Wess-Zumino-Witten (WZW) \cite{wzw} anomalous term, accounting for anomalous NGB processes, whose coefficient is fixed by topology arguments and is proportional to the number of colors $N_c$.

\subsection{Two flavors}

For simplicity, we will discuss the full
details of our approach in  the simpler case $N_f=2$.
Thus we will denote by the subscript $l$ the light
 quark correlator, and study $(\bar q q)_l\equiv\bar u u + \bar d d$ .
Note that we have defined the scalar source $s(x)$
as a matrix, but since for the physical partition function it corresponds to the mass matrix ${\cal M}$, which is diagonal, we are thus only interested in
the diagonal elements of $s(x)$ and  we can set the rest of the source terms to zero. In particular, for the two flavor case
${\cal M}=m\mathds{1}_2$ and we can write $s(x)=s_0(x)\mathds{1}_2$, so that:
\begin{eqnarray}
\condtwo_l &\equiv&
\frac{i}{Z_{QCD}[m]}\left.\frac{\delta Z_{QCD}[s_0]}{\delta s_0(x)}\right\vert_{s_0=m}\simeq \frac{i}{Z_{eff}[m]}\left.\frac{\delta Z_{eff}[s_0]}{\delta s_0(x)}\right\vert_{s_0=m}\equiv-\left\langle \frac{\delta {\cal L}_{eff}[s_0]}{\delta s_0(x)}\right\rangle_{\!\!\!\!s_0=m}.
\label{conddefsu2}
\end{eqnarray}
Proceeding in the same way now for the four light quark correlator:
\begin{eqnarray}
\langle T (\bar q q)_l(x)(\bar q q)_l(0)\rangle
&=&-\frac{1}{Z_{eff}[m]}\left.\frac{\delta}{\delta s_0(x)}\frac{\delta}{\delta s_0(0)}Z_{eff}[s_0]\right\vert_{s_0=m}\nonumber\\&=&-i\left\langle T\frac{\delta^2 {\cal L}_{eff}[s_0(x)]}{\delta s_0(x)^2}\right\rangle_{\!\!\!\!s_0=m}\!\!
\!\!\!\!\!\!\!\!\delta^{(D)}(x)+\left\langle T \frac{\delta {\cal L}_{eff}[s_0]}{\delta s_0(x)}\frac{\delta {\cal L}_{eff}[s_0]}{\delta s_0(0)}\right\rangle_{\!\!\!\!s_0=m}.
\label{Qdefsu2}
\end{eqnarray}
We will regularize all our expressions in dimensional regularization with  $D=4-\epsilon$ and for that purpose we keep the $D$-dependence in the $\delta$-function term above.

Now, from Eq.(\ref{Qdefsu2}),
and using the Lagrangians in  \cite{Gasser:1984gg,Bijnens:1999sh}, we obtain the following
result:
\begin{eqnarray}
\quarkcorl_{NLO}&=& 4B_0^2F^4\left\{1+\frac{4 M_{0\pi}^2}{F^2}\left(l^r_3+h^r_1\right) -6\mu_\pi  \right\} \label{condfoursu2nlo}\\
\quarkcorl_{NNLO}&=&\quarkcorl_{NLO}+4B_0^2F^4\left[\frac{2 M_{0\pi}^2}{F^2}\left(l^r_3+h^r_1\right) -3\mu_\pi  \right]^2\nonumber\\
&+&8B_0^2F^4\left[-\frac{3}{2}\mu_\pi^2 -\frac{3 M_{0\pi}^2}{F^2}\left(\mu_\pi\nu_\pi+4l^r_3\mu_\pi\right)+\frac{3M_{0\pi}^4}{8F^4}\left(-16l_3^r\nu_\pi+\hat c_1^{r}\right)\right],\nonumber\\
&+&B_0^2 \left[-8i(l_3+h_1)\delta^{(D)}(x)+ K^{(2)}(x)\right],
\label{condfoursu2nnlo}
\end{eqnarray}
where the NNLO constants $\hat{c}_i$ are defined in Eq.\eqref{l6mass} and, as usual \cite{Gasser:1984gg},
\begin{eqnarray}
\mu_\pi=\frac{M_{0\pi}^2}{32\pi^2 F^2}\log\frac{M_{0\pi}^2}{\mu^2}, \qquad
\nu_\pi=F^2\frac{\partial\mu_{0\pi}}{\partial M_{0\pi}^2}=\frac{1}{32\pi^2}\left(1+\log\frac{M_{0\pi}^2}{\mu^2}\right).
\label{nudefpi}
\end{eqnarray}
Note we have defined  $K^{(2)}(x)$ as the connected
part of the four-pion correlator to leading order:
\begin{equation}
K^{(2)}(x)=\langle T \phi^a (x)\phi_a (x) \phi^b (0)\phi_b (0)\rangle_{LO}-
\langle T \phi^a (0)\phi_a(0)\rangle^2_{LO}=6 \,G_\pi^2(x),
\label{Ksu2}
\end{equation}
 $G_\pi(x)$ being  the pion propagator to leading order
and the factor of $6=2(N_f^2-1)$ comes from the Wick contractions and
is nothing but twice the number of NGB fields.
The details of the renormalization and the dependence of the
constants $l_i^{r},h_i^r$ and $\hat c_i^{r}$ on the renormalization scale $\mu$
are given in Appendix \ref{app:renquark}.

\begin{figure}
\includegraphics[scale=0.55]{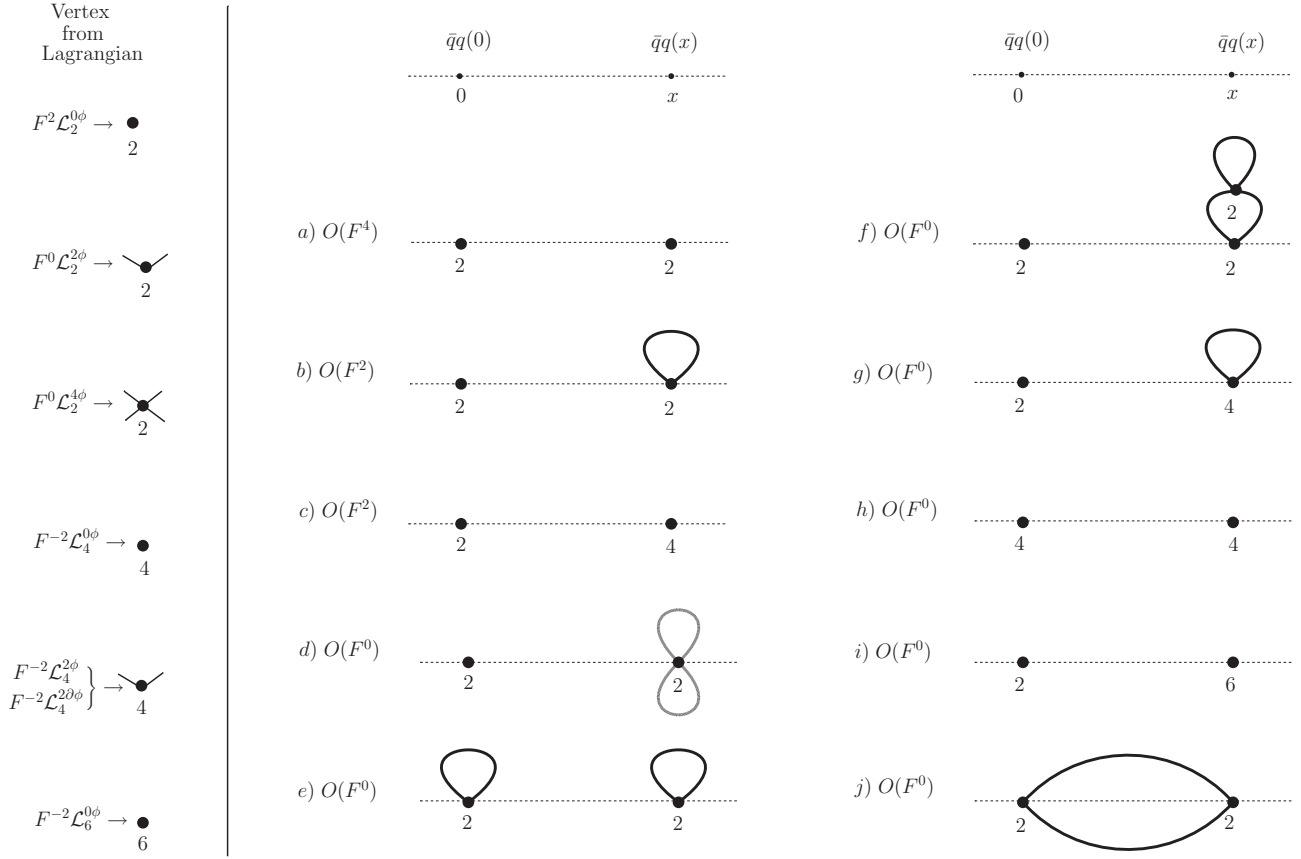}
  \caption{\rm \label{fig:diagfact} On the left column we provide the
    diagrammatic representation of the vertices coming from the different
    terms of the Lagrangian in Eq.\eqref{genformsu2}. The numbers attached to every
    vertex indicate the order of the Lagrangian. Diagrams a) to j) represent the
    different contributions to the four-quark correlator. The
    dotted horizontal line represents the spacetime separation between $0$ and
    $x$. Note that each NGB line decreases the order of the diagram by $1/F^2$. Diagram
    j)  is the first factorization-breaking term.  }
\end{figure}

To understand the structure of the different contributions to Eqs.\eqref{condfoursu2nlo} and \eqref{condfoursu2nnlo}
it is useful to recall the general form of the $SU(2)$ low-energy Lagrangian terms depending on the external scalar source.
For our NNLO calculation, we will need to keep terms up to $\Od(F^{-2})$.
 Let us then separate the terms
in the Lagrangian \cite{Gasser:1984gg,Bijnens:1999sh}, according to their $s$-dependence after expanding  the $U$  in NGB fields:
\begin{eqnarray}
{\cal L}_{eff}[s_0]&=&
\left({\cal L}_2^{0\phi} F^2+{\cal L}_2^{2\phi}+\frac{1}{F^2}{\cal L}_2^{4\phi}
+\frac{1}{F^2}{\cal L}_4^{2\partial\phi}\right) s_0
+\left({\cal L}_4^{0\phi}+\frac{1}{F^2}{\cal L}_4^{2\phi }\right)s_0^2\nonumber\\
&+&\frac{1}{F^2}{\cal L}_6^{0\phi} s_0^3+\frac{1}{F^2}\tilde{\cal L}_6^{0\phi}\partial_\mu s_0\partial^\mu s_0+\Od\left(\frac{1}{F^4}\right),
\label{genformsu2}
\end{eqnarray}
where we have also made explicit the leading $1/F^2$ dependence of each term.
The superscripts ``$n\phi$'' indicate the number of NGB fields or field derivatives on each Lagrangian contribution.
Note that, since ${\cal L}_k=\Od(p^k)$
in derivatives or $s$-powers ($s=\Od(p^2)$), it counts at least as
$\Od(1/F^{k-4})$, but the $1/F^2$ order of each term
grows when increasing the number of NGB fields, $\phi$.
We have represented the vertices arising from the different pieces of the Lagrangian above on the left column of Fig.\ref{fig:diagfact}.
 Note that all ${\cal L}_6$ terms in Eq.(\ref{genformsu2})
have the $0\phi$ superscript,
 because, to this order, they are simply constants. The constant
${\cal L}_6^{0\phi}$ term enters in $\condtwo_{l,NNLO}^2$
and ensures that one can renormalize the full result
so that the quark condensate is finite and scale-independent.
The term containing $(\partial s_0)^2$ does not contribute to this order.
The details as well as the explicit expression
of the condensates up to NNLO are given in  Appendix \ref{app:renquark}.

Once the structure of the vertices arising  from the
Lagrangian Eq.(\ref{genformsu2})
are understood,
we represent diagrammatically in Figure \ref{fig:diagfact}
the different contributions to $\quarkcor$.
On each diagram, the horizontal dotted line represents
spacetime, where each quark antiquark-bilinear stands at
separate points $0$ and $x$.
To LO and NLO -- respectively $\Od(F^4)$ and $\Od(F^2)$ --
all contributions are disconnected, as seen in diagrams a, b and c. The reason is that we can only use the
${\cal L}_2^{2\phi}$ term once and therefore, the NGB
line has to close upon itself -- a tadpole.
This gives  diagram (b) in Figure \ref{fig:diagfact}.
To NNLO ($\Od(F^0)$) we have all the possibilities shown in Figure
\ref{fig:diagfact} in diagrams (d)-(j). If one of the vertices comes from
${\cal L}_4$ or ${\cal L}_6$, once more there is at most one NGB line  and the
resulting diagram is disconnected. Note that among these is the
$\delta^{(D)}(x)$ term in Eq.(\ref{condfoursu2nnlo}) from diagram (h). With only
${\cal L}_2$ vertices, one has a diagram with a double tadpole
in one of the vertices
leading to a LO propagator squared at the same point (diagram (d)),
two vertices with one tadpole each (diagram (e)), a diagram like (b) but with
the propagator renormalized to NLO (diagram (f)) and another with two NGB
lines on each vertex but joined to form a connected one-loop diagram,
which is diagram (j). Actually, the latter is the only possible
connected contribution to this order, and gives the $G^2(x)$ term in Eq.(\ref{Ksu2}).
This whole discussion of vertices and diagrams will be
valid also for the $SU(3)$ case discussed below

Let us now turn to the factorization hypothesis and the relation between the four quark
correlation function and the two-quark condensate. We have collected in Appendix \ref{app:renquark}
all the two-quark condensates ChPT expressions up to NLO (given also in
\cite{Gasser:1983yg} for $SU(2)$ and in
\cite{Gasser:1984gg} for $SU(3)$) and up to NNLO, which have been given
explicitly in \cite{Amoros:2000mc} for $SU(3)$. Numerical estimations including NNLO corrections are given in \cite{Amoros:2000mc,Amoros:2001cp}.
In view of Eqs.(\ref{conddefsu2}),\eqref{condtwosu2nlo} and \eqref{condtwosu2nnlo}, it easy to check that
\begin{eqnarray}
&&\quarkcorl_{NLO}=\left(\condtwo_l^2\right)_{NLO},\nonumber\\
&&\quarkcorl_{NNLO}=\left(\condtwo_l^2\right)_{NNLO}+B_0^2 \left[-8i(l_3+h_1)\delta^{(D)}(x)+ K^{(2)}(x)\right].
  \label{resultssu2nnlo}
\end{eqnarray}

We see  that all  contributions  from disconnected diagrams in Figure \ref{fig:diagfact}, other than  the $\delta^{(D)}$ term,  can be absorbed in the two-quark condensate. Actually, up to NLO, we observe that $\quarkcorl$ in Eq.\eqref{condfoursu2nlo}
 is constant and equal to the NLO of the quark condensate squared, which
leads to factorization in the $N_c\rightarrow \infty$ limit
 (see section \ref{sec:fact} and \ref{sec:largen}).
However, to NNLO  the previous expression for $x=0$ contains the $G^2(0$)
divergent contribution, even after the quark condensate has been renormalized
and the $\delta^{(D)}$ term regularized.  We will show below that divergences cancel in physical quantities such as the scalar susceptibility, which is directly expressed in terms of observable quantities such as the free energy density. That is not the case for the four-quark condensate, which will remain divergent. Before analyzing these issues, let us extend the previous analysis to the  $SU(3)$ case.

\subsection{Three flavors}

In the $SU(3)$ case,  $\bar q q\equiv\bar u u +\bar d d + \bar s s$, ${\cal M}=\diag (m,m,m_s)$, $s(x)= \diag [s_0(x),s_0(x),s_s(x)]$ and:

\begin{eqnarray}
\condtwo &=&=-\left\langle \frac{\delta {\cal L}_{eff}[s_0]}{\delta s_0(x)}+\frac{\delta {\cal L}_{eff}[s_0,s_s]}{\delta s_s(x)}\right\rangle_{\!\!\!\!s={\cal M}},
\label{conddefsu3}
\end{eqnarray}

\begin{eqnarray}
\quarkcor&=&-i\left\langle T\left(\frac{\delta}{\delta s_0(x)}+\frac{\delta}{\delta s_s(x)}\right)^2{\cal L}_{eff}[s_0(x),s_s(x)]\right\rangle_{\!\!\!\!s={\cal M}}\!\!\!\!\!\!\!\!\delta^{(D)}(x)
\nonumber\\&+&\left\langle \left(\frac{\delta {\cal L}_{eff}[s_0,s_s] }{\delta s_0(x)}+\frac{\delta {\cal L}_{eff}[s_0,s_s]}{\delta s_s(x)}\right)\left(\frac{\delta {\cal L}_{eff}[s_0,s_s] }{\delta s_0(0)}+\frac{\delta {\cal L}_{eff}[s_0,s_s]}{\delta s_s(0)}\right)\right\rangle_{\!\!\!\!s={\cal M}}.
\label{Qdefsu3}
\end{eqnarray}
The $s$-dependent terms in the $SU(3)$ effective Lagrangian are now the
generalization of Eq.(\ref{genformsu2}) to include $s_s(x)$, so that we have
crossed terms like $s_0 s_s$, $s_0^2 s_s$
and so on, but the general structure is the same.
As in the $SU(2)$ case, the derivative terms  $(\partial s)^2$ do not
 contribute to $\quarkcor$ and thus
only four ${\cal L}_6$ constant terms contribute to renormalization.
As seen in  Appendix \ref{app:renquark}, they are
proportional to the $\hat{C}_i$ LEC given in Eq.\eqref{l6mass}.
Since we already presented the detailed discussion for the $SU(2)$ case in the previous section,
for the sake of brevity we cast our $SU(3)$ results for $\quarkcor$, which are much longer than before,  directly in terms of the two-quark condensates, namely,
\begin{eqnarray}
&&\quarkcor_{NLO}=\left(\condtwo^2\right)_{NLO},\nonumber\\
&&\quarkcor_{NNLO}=\left(\condtwo^2\right)_{NNLO}+B_0^2 \left[-24i(12L_6+2L_8+H_2)\delta^{(D)}(x)+ K(x)\right].
\label{resultssu3nnlo}
\end{eqnarray}
where $K(x)$ is the extension of Eq.(\ref{Ksu2}) to the $SU(3)$ case:
\begin{equation}
K(x)=\langle T \phi^a (x)\phi_a (x) \phi^b (0)\phi_b (0)\rangle_{LO}-\langle T \phi^a (0)\phi_a(0)\rangle^2_{LO}=2\left[3G_\pi^2(x)+4G_K^2(x)+G_\eta^2(x)\right].
\label{Ksu3}
\end{equation}
The ChPT expressions for the four-quark condensates
 to NNLO given in Eqs.(\ref{resultssu2nnlo}) and (\ref{resultssu3nnlo})
 (simplified in terms of the explicit expressions for $\condtwo_{NNLO}$, which are given in Appendix \ref{app:renquark}), are among the main results of the present work.

Note that, as it happened in the $SU(2)$ case, the contribution Eq.(\ref{Ksu3}) stems from
 $2(N_f^2-1)$ NGB propagators, although this time they have different masses.
Similarly, we can calculate separately the strange and non-strange four-quark condensates, which also factorize up to NLO, whereas to NNLO we get:
\begin{eqnarray}
\quarkcorl \!\!&=& \!\!\condtwo_{l,SU(3)}^2+B_0^2 \left[-16i(8L_6+2L_8+H_2)\delta^{(D)}(x)+6G_\pi^2(x)+2G_K^2(x)+\frac{2}{9}G_\eta^2(x)\right] +\Od\left(\frac{1}{F^2}\right),\qquad\label{Qlsu3}\\
\langle T(\bar s s)(x) (\bar s s)(0) \rangle \!\! &=& \!\!\langle \bar s s \rangle^2+B_0^2 \left[-8i(4L_6+2L_8+H_2)\delta^{(D)}(x)+2G_K^2(x)+\frac{8}{9}G_\eta^2(x)\right] +\Od\left(\frac{1}{F^2}\right).
\label{Qssu3}
\end{eqnarray}
Once again the explicit expressions for the renormalized
 $\condtwo_{NNLO}$ are given in Appendix \ref{app:renquark}.

The $\langle T(\bar q q)(x) (\bar s s)(0) \rangle$ correlator
has been calculated up to NNLO in \cite{Moussallam:2000zf,Bijnens:2006zp} in terms of the basis of the solutions to the
Muskhelishvili-Omn\`{e}s equations .

We remark that the four-quark correlators to NNLO given in Eqs.(\ref{resultssu2nnlo}) and (\ref{resultssu3nnlo}) are key ingredients to define the four-quark condensate and study the factorization hypothesis, as explained in the introduction.

\section{The scalar susceptibility}
\label{sec:sus}

In this section we will provide a consistency check of our calculation by
analyzing the chiral or scalar susceptibility to the first nontrivial order,
which can be obtained either differentiating the two-quark condensates or by
integration of the four-quark ones. The susceptibility is defined in Euclidean space-time as
\begin{equation}
\chi_l\equiv-\frac{\partial}{\partial m}\condtwo_l
\label{defsuscep}
\end{equation}
and measures the condensate thermal fluctuations, growing dramatically
near the chiral restoration, as confirmed by different lattice studies \cite{Karsch:2008ch}.
Therefore, let us consider the Euclidean (imaginary-time $t=-i\tau$) version of
 Eq.(\ref{zqcd}) and (\ref{zeff}), replacing $i\int d^4 x \rightarrow \int
 d\tau \int d^3 \vec{x}\equiv\int_E d^4x$  and the $(-,-,-,-)$ metric in the
 Lagrangian. Recall that the finite temperature $T$ case, which we will
 analyze elsewhere \cite{inpreptemp} would correspond to $\tau\in[0,\beta]$
 with $\beta=1/T$. In addition, in Eq.(\ref{Qdefsu2}) and (\ref{Qdefsu3}) we  have to replace  $-i\delta^D (x)\rightarrow \delta(\tau)\delta^{(D-1)}(\vec{x})\equiv \delta_E^D(x)$. With these replacements, we can now relate  the susceptibility with the four-quark correlators in the non-strange sector:

\begin{eqnarray}
\chi_l=\frac{1}{V_E}\frac{\partial^2}{\partial m^2}\log Z=\frac{1}{V_E}\left[\frac{1}{Z}\frac{\partial^2 Z}{\partial m^2}-\left(\frac{1}{Z}\frac{\partial Z}{\partial m}\right)^2\right]=\int_E d^D x \left[\quarkcorl-\condtwo_l^2\right].
\label{suslightdef}
\end{eqnarray}
where $V_E=\int_E d^D x$ is the $D$-dimensional Euclidean volume and $Z=Z[s={\cal M}]=e^{-zV_E}$ is the partition function with $z$ the free energy density.

The  relation in Eq.(\ref{suslightdef}) between $\chi_l$ and the four-point function  allows us to check our previous results. From Eqs.(\ref{resultssu2nnlo}) and  (\ref{Qlsu3}),  taking into account that:
\begin{equation}
\int_E d^D x \left[G_i (x)\right]^2=-\frac{d}{d M^2_i} G_i (0),
\end{equation}
and the expressions Eqs.(\ref{propreg})-(\ref{lambda}), together with the renormalization of the LEC in Eqs.(\ref{renelesgl}) and (\ref{gammaso4}), we obtain using the last integral in Eq.(\ref{suslightdef}):
\begin{eqnarray}
\chi_l^{SU(2)}&=&B_0^2\left[8\left(l^r_3(\mu)+h^r_1(\mu)\right)-12\nu_\pi\right]+\Od\left(\frac{1}{F^2}\right),\label{suscsu2}\\
\chi_l^{SU(3)}&=&B_0^2\left[16\left(8L_6^r(\mu)+2L_8^r(\mu)+H_2^r(\mu)\right)-12\nu_\pi-4\nu_K-\frac{4}{9}\nu_\eta\right]+\Od\left(\frac{1}{F^2}\right),
\label{suscsu3}
\end{eqnarray}
with $\nu_i$ given in Eq.(\ref{nudef}).

This is the same result that we get by taking directly the mass derivative of
the quark condensate to NLO in Eq.(\ref{condtwosu2nlo}) and (\ref{condtwosu3nlo})
using the leading order relations between meson and quark masses \cite{Gasser:1984gg}. This represents a  check of consistency of our calculation of the four-quark condensates to NNLO.
In addition, we have explicitly checked (using Eq.(\ref{renelesgl})) that the susceptibilities above are  finite and independent of the scale $\mu$. Furthermore,  with the conversion between the $SU(2)$ and $SU(3)$ LEC given in \cite{Gasser:1984gg}:
 \begin{equation}
l^r_3(\mu)+h^r_1(\mu)=2\left(8L_6^r(\mu)+2L_8^r(\mu)+H_2^r(\mu)-\frac{1}{4}\nu_K-\frac{1}{36}\nu_\eta,\right)
\end{equation}
we end up with:
$$\chi_l^{SU(2)}=\chi_l^{SU(3)}$$
 which is also consistent since the $SU(3)$ susceptibility is given by constant plus logarithmic terms in the $m_s\rightarrow\infty$ expansion, with no subleading terms in that expansion and therefore the very same expression has to be exactly recovered by calculating directly in the $SU(2)$ limit. Note also that the susceptibility to this order is independent of $F$.

Our result for the susceptibility  is also consistent  with a previous work
\cite{Smilga:1995qf} where only the leading infrared
order in the chiral limit was calculated,
namely the $\log M_{0\pi}^2$ term inside the $\nu_\pi$ in Eq.(\ref{suscsu2}). This is the expected behaviour of the susceptibility from the $O(4)$ model universality class near the chiral limit and below the critical temperature, namely $\chi\sim \log m$, with $m$
the mass of the non-strange quark \cite{Smilga:1995qf,Karsch:2008ch}.

We can follow the same procedure to obtain the strange quark susceptibility in terms of our strange four-quark correlation function:

\begin{eqnarray}
\chi_s\equiv-\frac{\partial}{\partial m_s}\conds=\frac{1}{V_E}\frac{\partial^2}{\partial m_s^2}\log Z=\frac{1}{V_E}\left[\frac{1}{Z}\frac{\partial^2 Z}{\partial m_s^2}-\left(\frac{1}{Z}\frac{\partial Z}{\partial m_s}\right)^2\right]=\int_E d^4 x \left[\langle T(\bar s s)(x) (\bar s s)(0) \rangle-\conds^2\right].
\label{susstrdef}
\end{eqnarray}
which, from Eq.(\ref{Qssu3}) gives:

\begin{equation}
\chi_s=B_0^2\left[8(4L_6^r+2L_8^r+H_2^r)-4\nu_K-\frac{16}{9}\nu_\eta\right].
\label{susstr}
\end{equation}
We have explicitly double checked this result
by taking the derivative
with respect to $m_s$ of the NLO strange quark condensate in Eq.(\ref{condsnlo}).
We remark that the results in Eqs.(\ref{suscsu2}), (\ref{suscsu3}) and (\ref{susstr}) for the ChPT scalar susceptibilities have not been given elsewhere.

\section{Non-Factorization}
\label{sec:fact}

 As explained in the introduction, we define the four quark condensate through
 Eq.(\ref{ourdef4q}), although in Appendix \ref{app:usualmsbar} we show that
 this is equivalent to the more usual definition of Eq.(\ref{usualdef4q}).  Therefore, by taking the $x\to 0$ limit in   Eqs.\eqref{resultssu2nnlo} and \eqref{resultssu3nnlo},
and despite $\delta^{(D)}(0)$ vanishes identically in dimensional
regularization \cite{Leibbrandt:1975dj} (now we are not integrating over $x$ as for the scalar susceptibility), there is still a term that
clearly breaks factorization, as defined in Eq.(\ref{facthyp}).
In particular,  we get in $SU(2)$, from Eq.(\ref{resultssu2nnlo}):
\begin{equation}
\frac{\condfour}{\condtwo^2}=1+\frac{6}{F^4}G_\pi^2(0)+\Od\left(\frac{1}{F^6}\right),
\label{factbreakingsu2}
\end{equation}
whereas in $SU(3)$ from Eq.(\ref{Ksu3}), we find:
\begin{equation}
\frac{\condfour}{\condtwo^2}=1+\frac{2}{F^4}\left[3G_\pi^2(0)+4G_K^2(0)+G_\eta^2(0)\right]+\Od\left(\frac{1}{F^6}\right),
\label{factbreakingsu3}
\end{equation}
where the propagators $G_i(0)$ are  given in dimensional regularization in Eq.(\ref{propreg}).

The non-factorization terms above are
divergent and independent of the LEC, once   the
two-quark condensate $\condtwo$ has been rendered finite  with the
renormalization of the $\Od(p^4)$ and $\Od(p^6)$
LEC  (see Appendix \ref{app:renquark}). The renormalizability of
$\condtwo$ is of course consistent with the fact that $ \bar q {\cal M} q$
is a QCD genuine observable RG-invariant.
Therefore, our non-factorization ChPT results in Eq.(\ref{factbreakingsu2}) and (\ref{factbreakingsu3}) imply
 that the
four-condensate  is divergent and
therefore the vacuum expectation value of $(\bar q q)^2$
does not admit a meaningful low-energy representation.

Our result is consistent with the one-loop QCD RG analysis in \cite{Narison:1983kn}, where  only one flavor is  considered.
In that paper it is shown that factorization is incompatible with the
renormalization group. Their argument goes as follows: the operator $(\bar q
q)^2$ mixes under renormalization with other four-quark operators, which can
be chosen in combinations such that their vacuum expectation values would
vanish if factorization holds. Then, assuming  factorization for those other
operators leads to the conclusion  that $\condfour$ is
divergent, which in particular means that it does not factorize
in terms of $\condtwo^2$ and that one cannot write any RG invariant made of four-quark operators.

Another interesting comment is that the
factorization breaking terms in Eqs.(\ref{factbreakingsu2}) and (\ref{factbreakingsu3}) vanish exactly
in the chiral limit, since
then
all dimensionally regularized propagators $G_\pi(0)=G_K(0)=G_\eta(0)=0$.
In that case, we would be forced to examine the neglected
NNNLO contributions in order to check the validity of factorization and the finiteness of the
 four-quark condensate. Recall that the arguments in \cite{Narison:1983kn} regarding four-quark operators hold actually for $m=0$.

\section{Large $N_c$}
\label{sec:largen}

Let us now discuss  the $N_f$ and $N_c$ dependence for the regularized
expression, namely, before taking the $D=4$ limit.
As we have checked for the $SU(2)$ and $SU(3)$ case, the connected four-field
functions in Eqs.(\ref{Ksu2}) and (\ref{Ksu3})
$K(x)=\Od(N_{GB})=\Od(N_f^2)$, where $N_{GB}=N_f^2-1$ is the number
of Goldstone Bosons. In addition, the $N_c$ leading behavior
of the different ChPT constants is well known \cite{Gasser:1984gg} from the QCD
$1/N_c$ expansion. In particular,
$F^2=\Od(N_c)$. Therefore, the first term that breaks factorization in
Eqs.(\ref{factbreakingsu2}) and (\ref{factbreakingsu3}) is $\Od\left(N_f^2/N_c^2\right)$,
which is rather different from the $1/(4N_fN_c)$ scaling
suggested in Eq.\eqref{facthyp}.
Unfortunately, we cannot say much more about the $N_f$
behavior of higher order terms, that could
change the global $N_f$ behavior. Note that the $N_f$
 dependence of the
quark correlators has been studied in detail in \cite{DescotesGenon:1999uh}
with a different motivation.

In the following, we will easily deduce
the $1/N_c$ behavior and, in particular,
we can study the large $N_c$ limit before
renormalization, and we will see that, in such formal case
factorization holds for $N_c\rightarrow\infty$.
First of all,
contrary to Eq.\eqref{facthyp},
in Eqs.\eqref{factbreakingsu2} and \eqref{factbreakingsu3}
there are no $\Od(1/N_c)$ terms. These could have
arisen from contributions of the type $L_i G(0)/F^4$, when
$L_i$ is $\Od(N_c)$, that actually appear in the
calculation. However, as we have said before, the whole
 $L_i$ dependence of the four-quark condensate is exactly
that of the two-quark condensate squared and thus
such terms do not break factorization.
The same happens with the $\Od(p^6)$ $c_i$ LEC in Eq.(\ref{l6mass}).
Still, one could wonder if $\Od(1/N_c)$ or larger $N_c$ powers could arise from
higher chiral orders that we have not calculated explicitly here.

Of course, as seen in Eqs.\eqref{factbreakingsu2} and \eqref{factbreakingsu3},
these higher chiral orders count at least as $\Od(1/F^6)$. Since $F^2=\Od(N_c)$, this already introduces a $1/N_c^3$ factor,
but it is not the only one, since the LEC can carry their own $N_c$ behavior. In particular,
we recall that, according to the chiral power counting discussed in sect.\ref{sec:fourq},
 the $\Od\left(1/F^n\right)$ contribution to the ratios in Eqs.(\ref{factbreakingsu2}) and (\ref{factbreakingsu3})
comes from connected diagrams with
  $n=2(L+1)+\sum_d N_d (d-2)$, with $L$ the number of loops and $N_d$
the number of vertices from ${\cal L}_d={\cal L}_2,{\cal L}_4,...$.  Note that
 a non-factorizing term requires at least
   $L=1$,  the leading contribution being the connected one-loop diagram (j) in Figure \ref{fig:diagfact} with two ${\cal L}_2$ vertices.
   This diagram yields the factorization breaking terms in Eqs.(\ref{factbreakingsu2}) and (\ref{factbreakingsu3}).

Now, the highest $N_c$ scaling of the LEC from ${\cal L}_d$ is $\Od(N_c^{(d-2)/2})$. The reason is that
these LEC, when divided by $F^{d-4}$ should yield
$\Od(N_c)$ contributions at  most, as
expected from the large-$N_c$ behaviour of the low-energy generating functional \cite{Gasser:1984gg}. This includes the WZW term, which is the anomalous part of
 ${\cal L}_4$ and is multiplied explicitly by $N_c$ \cite{wzw}. Although the WZW term does not depend on the quark mass, it could enter in this calculation through loop contributions. It is possible, of course, that some LEC do scale with a smaller $N_c$ power.
For instance, the $L_1$ to $L_{10}$ appearing in ${\cal L}_4$, are known to
scale as $\Od(N_c)$, except $L_4$, $L_6$ and $L_7$, which scale as $\Od(1)$.
These are model-independent QCD
predictions obtained in
 \cite{Gasser:1984gg}, with the exception of $L_7$,
which was taken there as $O(N_c^2)$.
This $L_7$ counting corresponds to integrating the $\eta'$ as a heavy particle
but then considering $m_{\eta'}^2\sim\Od(1/N_c)$ and therefore a light particle.
The consistent way of integrating the $\eta'$ yields
 $L_7\simeq\Od(1)$  \cite{Peris:1994dh}.
In summary, the $L_i$ in ${\cal L}_4$ are $\Od(N_c)$ at most, the $c_i$ in ${\cal L}_6$ are $\Od(N_c^2)$ at most, and
so on.

Hence, if a diagram has $N_d$ vertices from ${\cal L}_d$, they contribute, at most, with $N_d(d-2)/2$ powers of $N_c$.
Summing over all the $d$, the scaling of the LEC that contribute to that diagram
is given, at most, by $\sum_d  N_d(d-2)/2$ powers of $N_c$. Taking into account that the $1/F^n$ factors behave as $\Od(N_c^{-n/2})$,
we conclude that the non-factorization terms
should be $\Od(N_c^{\sum_d N_d (d-2)/2-(n/2)})=\Od(N_c^{-(L+1)})$ at most. But since we noted that
non-factorization terms require $L\geq1$, then
 the largest factorization breaking contribution is $\Od(N_c^{-2})$, at most.
Actually, this is the behavior of
the non-factorization correction we explicitly calculated
in Eqs.(\ref{factbreakingsu2}) and (\ref{factbreakingsu3}). This $\Od(N_c^{-2})$ counting of the factorization breaking, which we have
 formally showed here in the low-energy representation, confirms what had been suggested previously in the literature \cite{Ioffe:2005ym}.

Finally, if we compare with the original QCD factorization hypothesis
Eq.(\ref{facthyp}), we conclude that  factorization of the
four-quark condensate as the square of the two-quark condensate holds formally
in the  $N_c\rightarrow\infty$ limit. This is of course only a formal statement,
since we have just seen that in the low-energy calculation the factorization breaking terms diverge.

\section{Conclusions}
\label{sec:conclusions}

In this work we have addressed the issue of the four-quark condensate factorization into the two-quark condensate squared, within the low-energy representation
 of those condensates provided by Chiral Perturbation Theory.

  Our main result is the formal model-independent proof of the non-validity of the factorization or vacuum saturation hypothesis for the low-energy sector of QCD. A detailed calculation of the NNLO two-quark and four-quark condensates both for two and three flavors shows that, to that order, factorization is broken by terms which cannot be rendered finite with the usual renormalization procedure ensuring that the two-quark condensate is finite and scale-independent. This breaking of the factorization assumption at low-energies is then a model-independent result, since it relies only on the effective Lagrangian formalism, and is consistent with previous observations regarding the incompatibility of the factorization hypothesis with the QCD renormalization group evolution. In addition, the very same non-factorization term is obtained by using more conventional definitions of the quark condensate within the $\overline{MS}$ scheme in dimensional regularization.  As a consistency check of our analysis, we have derived the light and strange susceptibilities from the calculated four-quark correlators, showing that they agree with a direct derivative with respect to the quark masses of the two-quark condensates. The explicit renormalized and scale-independent expressions  for the ChPT NNLO susceptibilities are not given elsewhere, to our knowledge. Factorization would formally hold in the $N_c\rightarrow\infty$ limit, which we have been able to show to any order in the chiral expansion, the leading factorization breaking scaling as $\Od(1/N_c^2)$.

  We believe that these results can be useful for workers in the field, in particular concerning the OPE and sum-rule approach. A natural extension of this work is to consider finite temperature effects to see how they affect factorization and its connection with the chiral susceptibility, which in the thermal case plays a crucial role near chiral restoration \cite{inpreptemp}.

\appendix

\section{Quark condensates to NNLO in ChPT and their renormalization}
\label{app:renquark}

In this section we will give our NNLO results for the two-quark condensates.
As explained in text, the corresponding four-quark condensates
cannot  be obtained just by squaring these results, but one also has to add the non-factorizing contributions
described in Eqs.\eqref{resultssu2nnlo} and \eqref{resultssu3nnlo}.

The free meson propagator in dimensional regularization is given by \cite{Gasser:1984gg}:
\begin{equation}
G_i (0)=2M_{0i}^{D-2}\lambda,
\label{propreg}
\end{equation}
with
\begin{equation}
\lambda=\frac{\Gamma\left[1-\frac{D}{2}\right]}{2 (4\pi)^{D/2}},
\label{lambda}
\end{equation}
and $D=4-\epsilon$.

The $SU(3)$ ${\cal L}_4$ ChPT Lagrangian  is well known \cite{Gasser:1984gg} and we do not reproduce it here.
The relevant terms for the calculation of the condensates in the $\Od(p^6)$
Lagrangian \cite{Bijnens:1999sh} are only those dependent on the quark masses to leading order in the Goldstone boson fields. Here, we will follow, for simplicity, a different notation
than in \cite{Bijnens:1999sh} to denote  the ${\cal L}_6$ low-energy constants involved in the mass terms:
\begin{eqnarray}
{\cal L}_6^{m_q,SU(2)}
&=&\frac{B_0^3}{F^2}\hat c_1 m^3,\nonumber\\
{\cal L}_6^{m_q,SU(3)}&=&\frac{B_0^3}{F^2}\left(\hat C_1 m^3+\hat C_2m^2 m_s + \hat C_3 m
m_s^2 + \hat C_4 m_s^3\right)\label{l6mass},
\end{eqnarray}

Recall that our $\hat c_i$ are linear combinations of the LEC considered in \cite{Bijnens:1999hw,Bijnens:1999sh} whose precise form is not relevant here. Nevertheless, we  still  follow the convention in \cite{Bijnens:1999hw} for the renormalization of the $\Od(p^4)$ and  $\Od(p^6)$ LEC in the $\overline{\mbox{MS}}$ scheme:
\begin{eqnarray}
l_i&=&(c \mu)^{D-4}\left[l_i^r(\mu)+\gamma_i\Lambda\right],\nonumber\\
h_i&=&(c \mu)^{D-4}\left[h_i^r(\mu)+\delta_i\Lambda\right],\nonumber\\
\hat c_i&=&(c \mu)^{2(D-4)}\left[\hat c_i^r(\mu)-\hat\gamma_i^{(sq)}\Lambda^2-\left(\hat\gamma_i^{(0)}+\hat\gamma_i^{L}(\mu)\right)\Lambda\right]\nonumber\\
L_i&=&(c \mu)^{D-4}\left[L_i^r(\mu)+\Gamma_i\Lambda\right],\nonumber\\
H_i&=&(c \mu)^{D-4}\left[H_i^r(\mu)+\Gamma_i^H\Lambda\right],\nonumber\\
\hat C_i&=&(c \mu)^{2(D-4)}\left[\hat C_i^r(\mu)-\hat\Gamma_i^{(sq)}\Lambda^2-\left(\hat\Gamma_i^{(0)}+\hat\Gamma_i^{L}(\mu)\right)\Lambda\right], \label{bijrencon}
\end{eqnarray}
where $\mu$ is the renormalization scale, $\Lambda^{-1}=16\pi^2(D-4)$, $\log c=-\left[\log(4\pi)-\gamma+1\right]/2$, $\gamma=-\Gamma'[1]$, $\gamma_i$, $\delta_i$, $\Gamma_i$, $\Gamma_i^H$, $\hat\gamma_i^{(sq)}$, $\hat\Gamma_i^{(sq)}$, $\hat\gamma_i^{(0)}$ and $\hat\Gamma_i^{(0)}$   are numerical coefficients, whereas  $\hat\gamma_i^{L}$, $\hat\gamma_i^{L}$ are linear combinations of the $L_i^r(\mu)$. The above expression for the $\hat c_i$ shows that these constants have to absorb both two-loop divergences with ${\cal L}_2$ vertices and one-loop ones with one ${\cal L}_4$ and one ${\cal L}_2$ vertices.

The renormalization of the $L_i$ in Eq.(\ref{bijrencon}) coincides with that in \cite{Gasser:1984gg} up to $\Od(1)$ in the $\epsilon$ expansion:
\begin{equation}
L_i=L_i^r(\mu)+\Gamma_i\mu^{D-4}\lambda+\Od(\epsilon),
\label{renelesgl}
\end{equation}
and so on for the $H_i$, whereas the $l_i,h_i$ renormalization coincide with \cite{Gasser:1983yg} to that order.
For the renormalization of the one-loop effective action, the  $\Od(\epsilon)$ in Eqs.(\ref{bijrencon}) and (\ref{renelesgl}) can be neglected. However,
 when two-loop diagrams are considered, as it is our case here for the quark condensates (e.g., diagram (d) in Figure \ref{fig:diagfact}) products of the form $L_i G(0)$ yield finite contributions not vanishing in the $\epsilon\rightarrow 0^+$ limit. The $\Od(\epsilon)$ has to be kept also in the expansion of $\lambda$ in Eq.(\ref{lambda}) when expanding $G_i(0)$ in Eq.(\ref{propreg}) in $G_i(0)^2$ contributions.

As for the $\mu$ scale dependence,  the $L_i, l_i$ and the $\hat C_i, \hat c_i$ are scale-independent so that the scale dependence of the $L_i^r(\mu),l_i^r(\mu)$, $\hat C_i^r(\mu), \hat c_i^r(\mu)$ is canceled
 with the explicit $\mu$-dependence appearing in Eq.(\ref{bijrencon}). This allows to express all the logarithms of the masses referred to the scale $\mu$, i.e., $\log(M_i^2/\mu^2)$ so that the final result for the observables should be finite and scale-independent.

   We also recall that to the order we are calculating,
the  propagators are renormalized to NLO (tadpole corrections) and
one has to include the wave-function and mass renormalization to that order. The renormalized masses are given in
 \cite{Gasser:1984gg}, while the explicit wave-function renormalization
 can be found for instance in \cite{GomezNicola:2001as}. We recall that we should include
  now up to $\Od(\epsilon)$ in those tadpole corrections, for the reasons just explained.

With these renormalization conventions, we turn to the NNLO quark condensates. The $\Gamma_i$ coefficients appearing in the calculation are
 \cite{Gasser:1984gg}:
 \begin{eqnarray}
  \gamma_3&=&-1/2, \quad  \delta_1=2,\nonumber\\
 \Gamma_4&=&1/8, \quad  \Gamma_5=3/8, \quad \Gamma_6=11/144, \quad \Gamma_7=0, \quad \Gamma_8=5/48, \quad \Gamma_2^{H}=5/24,
 \label{gammaso4}
 \end{eqnarray}

 Recall that in $SU(3)$ $L_6$, $L_8$ and $H_2$ come explicitly from the ${\cal L}_4$ vertex contributions to the condensate and are therefore the only LEC appearing to NLO. The mass and wave function renormalization bring up also $L_4$, $L_5$ and $L_7$ to the final result. $L_7$ only appears in the $\eta$ mass renormalization. In the pure $SU(2)$ case,  only $l_3$ and $h_1$ enter in the calculation.

 Once the above LEC renormalization is performed, we have checked that one
 can choose the $\hat{c}_i$ and $\hat{C}_i$ in Eq.(\ref{l6mass}), renormalized through
 Eq.(\ref{bijrencon}),  so that the final result for the two-quark condensates is finite and scale independent.  We obtain:
 \begin{eqnarray}
\hat\gamma_1^{(sq)}&=&12,\nonumber\\
\hat\gamma_1^{L}&=&-48l_3^r,\nonumber\\
\hat\Gamma_1^{(sq)}&=&896/81, \quad \hat\Gamma_2^{(sq)}=32/27, \quad \hat\Gamma_3^{(sq)}=64/9, \quad \hat\Gamma_4^{(sq)}=160/81,  \nonumber\\
\hat\Gamma_1^{L}&=&\frac{32}{27} \left(444 L_4^r + 191 L_5^r - 6 (148 L_6^r + 4 L_7^r + 65 L_8^r)\right),\nonumber\\
\hat\Gamma_2^{L}&=&\frac{32}{9}  \left(162 L_4^r + 31 L_5^r - 324 L_6^r - 62 L_8^r\right),\nonumber\\
\hat\Gamma_3^{L}&=&\frac{32}{9} \left(96 L_4^r + 35 L_5^r - 192 L_6^r + 24 L_7^r - 62 L_8^r\right),\nonumber\\
\hat\Gamma_4^{L}&=&\frac{32}{27} \left(78 L_4^r + 43 L_5^r - 6 (26 L_6^r + 8 L_7^r + 17 L_8^r)\right),
  \end{eqnarray}
 and all the linear terms $\hat\gamma_i^{(0)}=\hat\Gamma_i^{(0)}=0$ for  the above LEC.

For convenience and following the same notation as \cite{Gasser:1984gg}, we  define:
 \begin{eqnarray}
 \mu_i=\frac{M_{0i}^2}{32\pi^2 F^2}\log\frac{M_{0i}^2}{\mu^2}, \qquad
 \nu_i=F^2\frac{\partial\mu_i}{\partial M_{0i}^2}=\frac{1}{32\pi^2}\left(1+\log\frac{M_{0i}^2}{\mu^2}\right).
 \label{nudef}
 \end{eqnarray}

In SU(2) the leading order pion mass is related with the physical one by:
\begin{equation}\label{physicalmassesSU2}
M_\pi^2=M_{0\pi}^2\left(1+\mu_{\pi}+\frac{4M_{0\pi}^2}{F^2}l_3^r\right),
\end{equation}
and in SU(3),
\begin{equation}\label{physicalmassesSU3}
\begin{split}
M_\pi^2=&M_{0\pi}^2\left[1+\mu_{\pi}-\frac{\mu_{\eta}}{3}+\frac{16M_{0K}^2}{F^2}(2L_6^r-L_4^r)+\frac{8M_{0\pi}^2}{F^2}\left(2L^r_6+2L^r_8-L_4^r-L_5^r\right)\right],\\
M_K^2=&M_{0K}^2\left[1+\frac{2\mu_\eta}{3}+\frac{8M_{0\pi}^2}{F^2}\left(2L_6^r-L_4^r\right)+\frac{8M_{0K}^2}{F^2}\left(4L_6^r+2L_8^r-2L_4^r-L_5^2\right)\right],\\
M_\eta^2=&M_{0\eta}^2\left[1+2\mu_K-\frac{4}{3}\mu_{\eta}+\frac{8M_{0\eta}^2}{F^2}\left(2L_8^r-L_5^r\right)+\frac{8}{F^2}\left(2M_{0K}^2+M_{0\pi}^2\right)\left(2L_6^r-L_4^r\right)\right]+\\
&M_{0\pi}^2\left[-\mu_\pi+\frac{2}{3}\mu_K+\frac{1}{3}\mu_{\eta}\right]+\frac{128}{9F^2}\left(M_{0K}^2-M_{0\pi}^2\right)^2\left(3L^r_7+L_8^r\right).
\end{split}
\end{equation}

The relation between the leading order pion decay constant and the physical
one up to two-loops in given in \cite{Bijnens:1997vq} for SU(2) and in
\cite{Amoros:1999dp} for SU(3).

The final expressions for the  two-quark condensates, finite and
scale-independent, up to NNLO,  that have been calculated previously in \cite{Amoros:2000mc} for $SU(3)$, are given by:
\begin{eqnarray}
\condtwo^{SU(2)}_{l,NLO}&=& -2B_0F^2\left\{1+\frac{2 M_{0\pi}^2}{F^2}\left(h_1^r+l_3^r\right) -3\mu_\pi  \right\} \label{condtwosu2nlo}\\
\condtwo^{SU(2)}_{l,NNLO}&=&\condtwo^{SU(2)}_{l,NLO}-2B_0F^2\left[
-\frac{3}{2}\mu_\pi^2 -\frac{3 M_{0\pi}^2}{F^2}\left(\mu_\pi\nu_\pi+4l_3^r\mu_\pi\right)\right.\nonumber\\
&+&\left.\frac{3M_{0\pi}^4}{8F^4}\left(-16l^r_3\nu_\pi+\hat c_1^r\right)\right],
\label{condtwosu2nnlo}
\end{eqnarray}

\begin{eqnarray}
\condtwo^{SU(3)}_{l,NLO}&=& -2B_0F^2\left\{1+\frac{4}{F^2}\left[\left(H_2^r + 4 L_6^r + 2 L_8^r\right) M_{0\pi}^2+8 L_6^r M_{0K}^2  \right]-3\mu_\pi -2\mu_K-\frac{1}{3}\mu_\eta \right\}\label{condtwosu3nlo}\\
\condtwo^{SU(3)}_{l,NNLO}&=&\condtwo^{SU(3)}_{l,NLO}-2B_0F^2\left\{
-\frac{3}{2}\mu_\pi^2 +\frac{1}{18}\mu_\eta^2+\mu_\pi\mu_\eta-\frac{4}{3}\mu_K\mu_\eta
+\frac{1}{F^2}\left[-3M_{0\pi}^2\mu_\pi\nu_\pi+\frac{1}{3}M_{0\pi}^2\mu_\pi\nu_\eta\right.\right.\nonumber\\
&-&\left.\frac{8}{9}M_{0K}^2\mu_K\nu_\eta+M_{0\pi}^2\mu_\eta\nu_\pi-\frac{4}{3}M_{0K}^2\mu_\eta\nu_K
+\frac{1}{27}\left(16M_{0K}^2-7M_{0\pi}^2\right)\mu_\eta\nu_\eta\right]\nonumber\\
&+&\frac{24}{F^2}\mu_\pi\left[\left(3 L_4^r + 2 L_5^r - 6 L_6^r - 4 L_8^r\right)M_{0\pi}^2+2\left(L_4^r - 2 L_6^r\right)M_{0K}^2\right]
\nonumber\\&+&\frac{16}{F^2}\mu_K\left[\left(L_4^r - 2 L_6^r\right)M_{0\pi}^2+2\left(3 L_4^r +  L_5^r - 6 L_6^r - 2 L_8^r\right)M_{0K}^2\right]\nonumber\\
&+&\frac{8}{9F^2}\mu_\eta\left[\left(-3 L_4^r - 2 L_5^r + 6 L_6^r - 48 L_7^r - 12 L_8^r\right)M_{0\pi}^2+2\left(15 L_4^r + 4 L_5^r - 30 L_6^r + 24 L_7^r\right)M_{0K}^2\right]\nonumber\\
&+&\frac{24M_{0\pi}^2}{F^4}\nu_\pi\left[\left(L_4^r + L_5^r - 2 L_6^r - 2 L_8^r\right)M_{0\pi}^2+2\left(L_4^r - 2 L_6^r\right)M_{0K}^2\right]\nonumber\\
&+&\frac{16M_{0K}^2}{F^4}\nu_K\left[\left(L_4^r - 2 L_6^r\right)M_{0\pi}^2+\left(2 L_4^r + L_5^r - 4 L_6^r - 2 L_8^r\right)M_{0K}^2\right]\nonumber\\
&+&\frac{8}{27F^4}\nu_\eta\left[\left(-3 L_4^r + L_5^r + 6 L_6^r - 48 L_7^r - 18 L_8^r\right)M_{0\pi}^4+2\left(3 L_4^r - 4 L_5^r - 6 L_6^r + 48 L_7^r + 24 L_8^r\right)M_{0\pi}^2 M_{0K}^2\right.\nonumber\\&+&\left.8\left(3 L_4^r + 2 (L_5^r - 3 (L_6^r + L_7^r + L_8^r))\right)M_{0K}^4\right]\nonumber\\
&+&\left.\frac{1}{8F^4}\left[\left(3 \hat C_1^r - 2 \hat C_2^r +\hat
      C_3^r\right)M_{0\pi}^4+4\left(\hat C_2^r - \hat C_3^r\right)M_{0\pi}^2 M_{0K}^2+4\hat C_3^rM_{0K}^4\right]
\right\},
\label{condtwosu3nnlo}
\end{eqnarray}

\begin{eqnarray}
\langle \bar s s \rangle_{NLO}&=&-B_0F^2\left\{1+\frac{4}{F^2}\left[-\left(H_2^r - 4 L_6^r + 2 L_8^r\right) M_{0\pi}^2+2\left(H_2^r + 4 L_6^r + 2 L_8^r\right) M_{0K}^2  \right]-4\mu_K-\frac{4}{3}\mu_\eta \right\}\label{condsnlo}\\
\langle \bar s s \rangle_{NNLO}&=&\langle \bar s s \rangle_{NLO}-B_0F^2\left\{
\frac{8}{9}\mu_\eta^2-\frac{8}{3}\mu_K\mu_\eta
+\frac{1}{F^2}\left[\frac{4}{3}M_{0\pi}^2\mu_\pi\nu_\eta\right.\right.\nonumber\\
&-&\left.\frac{32}{9}M_{0K}^2\mu_K\nu_\eta-\frac{8}{3}M_{0K}^2\mu_\eta\nu_K
+\frac{4}{27}\left(16M_{0K}^2-7M_{0\pi}^2\right)\mu_\eta\nu_\eta\right]\nonumber\\
&+&\frac{48}{F^2}\mu_\pi\left(L_4^r - 2 L_6^r\right)M_{0\pi}^2
\nonumber\\&+&\frac{32}{F^2}\mu_K\left[\left(L_4^r - 2 L_6^r\right)M_{0\pi}^2+2\left(2 L_4^r +  L_5^r - 4 L_6^r - 2 L_8^r\right)M_{0K}^2\right]\nonumber\\
&+&\frac{16}{9F^2}\mu_\eta\left[\left(3 L_4^r - 4 L_5^r - 6 L_6^r + 48 L_7^r + 24 L_8^r\right)M_{0\pi}^2+8\left(3 L_4^r + 2 (L_5^r - 3 (L_6^r + L_7^r + L_8^r))\right)M_{0K}^2\right]\nonumber\\
&+&\frac{32M_{0K}^2}{F^4}\nu_K\left[\left(L_4^r - 2 L_6^r\right)M_{0\pi}^2+\left(2 L_4^r + L_5^r - 4 L_6^r - 2 L_8^r\right)M_{0K}^2\right]\nonumber\\
&+&\frac{32}{27F^4}\nu_\eta\left[\left(-3 L_4^r + L_5^r + 6 L_6^r - 48 L_7^r - 18 L_8^r\right)M_{0\pi}^4+2\left(3 L_4^r - 4 L_5^r - 6 L_6^r + 48 L_7^r + 24 L_8^r\right)M_{0\pi}^2 M_{0K}^2\right.\nonumber\\&+&\left.8\left(3 L_4^r + 2 (L_5^r - 3 (L_6^r + L_7^r + L_8^r))\right)M_{0K}^4\right]\nonumber\\
&+&\left.\frac{1}{4F^4}\left[\left(\hat C_2^r - 2 \hat C_3^r + 3 \hat C_4^r\right)M_{0\pi}^4+4\left(\hat C_3^r - 3 C_4^r\right)M_{0\pi}^2 M_{0K}^2+12\hat C_4^r M_{0K}^4\right]\right\},
\label{condsnnlo}
\end{eqnarray}
where  the Gell-Mann-Okubo relation $3M_{0\eta}^2=4M_{0K}^2-M_{0\pi}^2$ for
the $SU(3)$ leading order masses has been used and the renormalized $L_i^r,l_i^r$ and $\hat c_i^r,\hat C_i^r$ constants depend on the scale $\mu$ as explained above.

\section{Four-quark condensates in the usual $\overline{MS}$ definition}
\label{app:usualmsbar}

Here we consider the definition in Eq.(\ref{usualdef4q}) of the four-quark condensate in Euclidean space. Let us restrict to $SU(2)$ since it will become clear that the argument can be straightforwardly extended to the $SU(3)$ case. The four-quark correlator to NNLO is given in Eq.(\ref{resultssu2nnlo}), so that its Euclidean Fourier transform to this order is (see our Euclidean space-time conventions in section \ref{sec:sus}):

\begin{equation}
\Pi (Q^2)=(2\pi)^D\condtwo^2\delta^{(D)}(Q) + 2B_0^2\left[4(l_3+h_1)+3J_\pi(Q^2)\right]
\label{PiQ}
\end{equation}
with $Q^2=\sum_{i=1}^D Q_i^2$ and

\begin{equation}
J_\pi(Q^2)=\int\frac{d^D K}{(2\pi)^D} G_\pi (K) G_\pi (K-Q),
\label{Jdef}
\end{equation}
which is nothing but the one-loop integral appearing in pion-pion scattering,
dimensionally regularized in \cite{Gasser:1983yg}. Its divergent part is
contained in $J_\pi(0)=-2M_\pi^{D-4}\lambda-1/(16\pi^2)$ with $\lambda$
defined in Eq.(\ref{lambda}) while $\bar{J} (Q^2)=J_\pi(Q^2)-J_\pi(0)$ is
finite. Note also that $J_\pi(Q^2)$ defined in Euclidean space is real. The
imaginary part in $\bar J_\pi$ giving the usual unitarity cut in scattering
amplitudes arises when the analytical continuation of $Q^2$ to Minkowski
space-time is performed, but here we should keep the Euclidean version, since
we are following the prescription in Eq.(\ref{usualdef4q}) to perform the additional momentum integral.

Before proceeding to the calculation of the four-quark condensate, let us note that the divergent part of the $J_\pi$ in Eq.(\ref{PiQ}) cancels exactly with the LEC contribution since  $l_3+h_1=l_3^r(\mu)+h_1^r(\mu)+(3/2)\mu^{D-4}\lambda$ (see Eqs.(\ref{bijrencon}) and (\ref{gammaso4})). Thus, $\Pi(Q^2)$ is finite and scale-independent before integration in $Q$. This is actually a welcomed check, since the scalar susceptibility given in Eq.(\ref{suslightdef}) can be written also as $\chi_l=\tilde\Pi (0)$ with $\tilde\Pi(Q^2)=\Pi(Q^2)-(2\pi)^D\condtwo^2\delta^{(D)}(Q)$ and should be finite and scale-independent.

However, we will immediately see that the additional integration in $Q$ in
Eq.(\ref{usualdef4q}) generates an extra divergence which cannot be removed
and in the end gives the same divergent factorization-breaking result as the
definition in Eq.(\ref{ourdef4q}). For that purpose, let us follow the standard dimensional regularization procedure \cite{Leibbrandt:1975dj} and write:

\begin{equation}
J_\pi(Q^2)=\frac{1}{(4\pi)^{D/2}}\int_0^1 dx \int_0^\infty d\lambda \lambda^{1-D/2}\exp\left\{ -\lambda \left[M_\pi^2+Q^2 x(1-x)\right]\right\}
\end{equation}
which is valid within the domain $\re[D]<4$. Now, before performing the $x$ and $\lambda$ integrals above, we integrate over $Q$ so that:

\begin{equation}
\int\frac{d^D Q}{(2\pi)^D} J_\pi(Q^2)=\frac{1}{(4\pi)^D}\left\{\int_0^1 dx \left[x(1-x)\right]^{-D/2}\right\}\left\{\int_0^\lambda d\lambda \lambda^{1-D} e^{-\lambda M_\pi^2}\right\}=\frac{\left(M_\pi^2\right)^{D-2}}{(4\pi)^D}\left[\Gamma\left(1-\frac{D}{2}\right)\right]^2=G_\pi^2(0)
\end{equation}
 where the one-dimensional integrals are solved for $\re[D]<2$ and we have
 used standard properties of the Gamma function. Since the result is analytic
 in $D$, it can be extended to $D=4-\epsilon$ with $\epsilon\rightarrow
 0^+$. Therefore, integrating in Eq.(\ref{PiQ}) over $Q$ according to
 Eq.(\ref{usualdef4q}) and taking into account that $\int d^D
 Q/(2\pi)^D=\delta^{(D)}(0)=0$,  gives exactly the same divergent
 factorization-breaking result for the four-quark condensate as the one using
 the prescription of Eq.(\ref{ourdef4q}).

Another way to arrive to the same conclusion is to perform the change of variables $Q\rightarrow Q+K$ in the double $D$-integral $\int d^D Q \int d^D K$ in the region of $D$ where it converges, which in this case is $\re[D]<2$, which follows by direct power counting in $Q$ and $K$ of the propagators in Eq.(\ref{Jdef}) in the large $Q^2$ and $K^2$ Euclidean region.

It is clear that the same equivalence between the two definitions holds in the $SU(3)$ case simply by considering  $J_K$ and $J_\eta$ apart from $J_\pi$, since the results of the correlators in Eqs.(\ref{Qlsu3}) and (\ref{Qssu3}) do not mix different meson species.

\section*{Acknowledgments}
Work partially supported by Spanish Ministerio de
Educaci\'on y Ciencia research contracts: FPA2008-00592, FIS2006-03438, FIS2008-01323 and
U.Complutense/Banco Santander grant UCM-BSCH GR58/08 910309. We acknowledge the support
of the European Community-Research Infrastructure
Integrating Activity
``Study of Strongly Interacting Matter" 
(acronym HadronPhysics2, Grant Agreement
n. 227431)
under the Seventh Framework Programme of EU.

\end{document}